\documentclass[conference]{IEEEtran}

\usepackage[]{graphicx,color}
\usepackage{amsmath,amssymb}
\usepackage{bm}
\usepackage{ascmac}
\usepackage{url}                      
\usepackage{hyperref}
\usepackage{ifthen}
\usepackage[T1]{fontenc}

\IEEEsettopmargin{t}{0.71in}
\newcommand{\F}[0]{{\mathbb F}}

\newcommand{\argmax}{\mathop{\rm arg~max}\limits}

\begin{document}

\title{Compute-and-forward {relaying} with LDPC codes
over QPSK scheme
}

\author{
 \IEEEauthorblockN{
Satoshi Takabe\IEEEauthorrefmark{1},
Tadashi Wadayama\IEEEauthorrefmark{1}, 
\'{A}ngeles Vazquez-Castro\IEEEauthorrefmark{2}, and                        
Masahito Hayashi\IEEEauthorrefmark{3}\IEEEauthorrefmark{4}\IEEEauthorrefmark{5}}
  \IEEEauthorblockA{\IEEEauthorrefmark{1}
  	Department of Computer Science, Faculty of Engineering, Nagoya Institute of Technology}
  \IEEEauthorblockA{\IEEEauthorrefmark{2}
  	 Department of Telecommunications and Systems Engineering, Autonomous University of Barcelona}
  \IEEEauthorblockA{\IEEEauthorrefmark{3}
  	 Graduate School of Mathematics, Nagoya University}
  \IEEEauthorblockA{\IEEEauthorrefmark{4}
  	 Shenzhen Institute for Quantum Science and Engineering, Southern University of Science and Technology}
  \IEEEauthorblockA{\IEEEauthorrefmark{5}
  	 Centre for Quantum Technologies, National University of Singapore}
  \IEEEauthorblockA{
  	Email: \{s\_takabe, wadayama\}@nitech.ac.jp, angeles.vazquez@uab.cat,
  	masahito@math.nagoya-u.ac.jp}
                    } 
\maketitle


\begin{abstract}
In this paper, we study a compute-and-forward (CAF) relaying scheme with low-density parity-check (LDPC) codes, a special case of physical layer network coding, under the quadrature phase shift keying (QPSK) modulation. The novelty of this paper is the theoretical analysis of decoding performance of the CAF scheme and traditional separation decoding (SD) scheme with joint decoding or with successive interference cancellation (SIC) decoding when the reception powers from both senders are not equal but close to each other. First, we study the decoding performance of linear random coding (LRC) in the CAF scheme whose decoder is based on the degraded channel. When rotation angles of constellations of two senders are controlled, we show that they can achieve rates well beyond the multiple access channel (MAC) with LRC with optimal rotation angles. Second, we analyze the practical feasibility of CAF using LDPC codes in terms of computational costs and decoding performance of belief propagation (BP) decoders. The calculation complexity of the BP decoder for the CAF scheme is almost equal to that of the SIC BP decoder, and smaller than the joint BP decoder in the SD scheme.  Decoding performance is evaluated by the asymptotic decodable region using density evolution. The results show that, with code rate fixed, the CAF scheme has better performance than the SD scheme with joint BP decoding and SIC BP decoding in the high rate region. 
\end{abstract}

\section{Introduction}
The advent of 5G wireless networks requires the development of novel access techniques to enable a massive increase in the number of users. Non-orthogonal multiple access (NOMA) is recognised as an enabling technology as it allows simultaneous access of users in time, frequency, and waveform thus naturally increasing the number of active users in a wireless network. 
The most widely proposed NOMA so far exploits the power domain and uses successive interference cancellation (SIC) with low-density parity-check (LDPC) codes \cite{Miridakis} because the maximum likelihood (ML) decoder 
with SIC achieves the multiple access (MAC) capacity region corners \cite{Palanki}. 
Performance of SIC decoding is poor when users have comparable received SNRs. 
For this reason, strong and weak users throughout the coverage need first to be clustered and paired to ensure fair rate allocation \cite{Rimoldi}. This feature makes the technique not fully optimal for general deployment since fair clustering and pairing of strong and weak users is not guaranteed. 
This is specially true in our scenarios of interest, {e.g.,} integrated wireless networks (see Fig. 1) where there may be subnetworks with balanced or unbalanced SNRs within and/or across the several subnetworks' coverage. 
Furthermore, traditional {separation decoding (SD)} 
is complex and not easy to implement \cite{Yedla}.
As a result, we envision different NOMA techniques coordinately used in integrated wireless networks. 

\begin{figure}[t]
    \centering
    \includegraphics[scale=0.3]{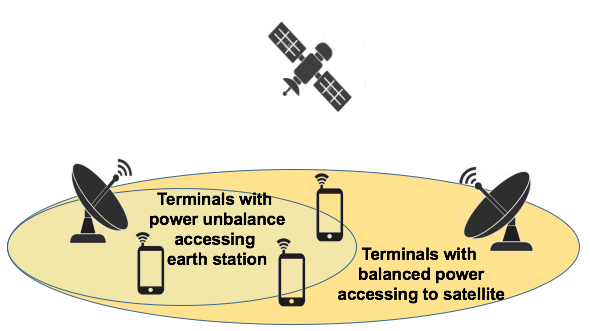}
    \caption{Uplink NOMA scenario with integrated wireless systems conveying terminals of different capabilities and subnetworks with homogeneous or heterogeneous SNRs.}
    \label{scenario}
\end{figure}
\begin{figure}[t]
    \centering
    \includegraphics[scale=0.37]{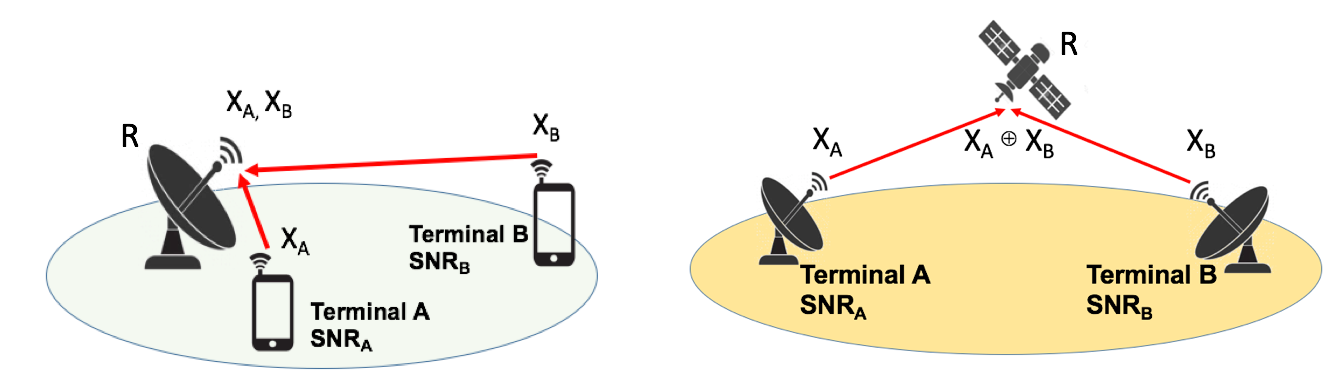}
    \caption{Uplink NOMA signals in the integrated wireless system with illustrative cases of SD (left) and CAF (right) schemes.}
    \label{scenario}
\end{figure}

The NOMA technique {that} we propose exploits physical layer network coding (PNC), a known capacity boosting concept as proposed in \cite{Nazer11}, which we term as \emph{compute-and-forward (CAF) scheme}. In CAF, the receiver tries to {\em decode} the modulo sum of the codewords sent from the transmitting terminals. In Fig. 2, we show a simplified model called 
\emph{two-way relay channel} of an integrated wireless scenario in Fig. 1 for the two-user case. In this case, the terrestrial access point and the satellite act as relays (R) and terminals A and B are the terminals accessing the channel in the uplink. On the left, the receiver performs SD scheme while on the right the receiver {employs the CAF scheme}.  
 The same technique has already been proposed in \cite{Lu14}. This work combines the joint use of physical-layer network coding and practical multiuser decoding for unbalanced SNRs scenarios. The work in \cite{Guo2018} considers the same problem and proposes a different decoding technique based on \cite{Nazer2016} that improves outage and fairness. 

The study \cite{Nazer11} is based on lattice codes, whose practical implementation is harder than conventional linear codes over $\F_2$. While linear codes is studied in~\cite{Lu14}, it employs Reed-Solomon code, whose performance is not so high.


{Unlike those studies,} for the practical feasibility of CAF, we employ 
LDPC codes
because a combination of LDPC codes and belief propagation (BP) decoder has been proved to be very powerful and effective for additive noise channels \cite{MacKay99}. 
Sula {\it et al.} \cite{Sula17} discussed regular LDPC codes with a modified BP decoding for CAF.
Some of the authors \cite{Takabe18} studied how spatial-coupled LDPC codes improved the performance for CAF.
Its BP decoding has almost similar calculation complexity as
SIC BP decoding of SIC in the SD scheme.
These studies for LDPC codes were done through asymptotic analysis of {\em density evolution} (DE) \cite{Richardson}.

{It should be emphasized that} the preceding studies \cite{Sula17,Takabe18}
 address only the case when the powers of the signals from both senders are the same
while it is difficult to control the powers of both signals to satisfy this condition.
In addition, they assumed the binary phase shift keying (BPSK) as modulation,
 {and treated only real-valued signals. 
 For example, when there is only one sender, 
the {quadrature phase shift keying (QPSK)} modulation can be reduced to twice transmission in the BPSK scheme.
On the other hand, when there are two senders, this reduction is not always possible 
because the difference $\theta$ of rotation angles between the constellations of two senders is not zero in general.
We thus need to analyze the CAF scheme on the complex plane if we consider a higher-order modulation.}

In this paper, to optimize the angle $\theta$,
under the SD and CAF scheme,
we focus on the mutual information rate with the uniform prior distribution,
which is called symmetric information rate (SIR)
because it expresses the transmission rate using linear random coding (LRC).
Optimizing the SIR as the function of this angle,
we clarify that 
the angle $\theta=0$
realizes the optimal SIR in the CAF scheme and that
the angle $\theta=\pi/4$ realizes almost the optimal SIR in the SD scheme.

As the next step, we explore the practical feasibility and implementation of CAF using regular LDPC codes
 when $\theta=0$
but the signal powers from two senders A and B are not the same.
Then, we compare theoretical performances of the CAF scheme (based on degraded channel as in \cite{Ullah17}) and the SD scheme for LRC and ML decoder.
Also, we compare our method with 
the SD scheme of regular LDPC codes and a joint BP decoder when $\theta=0$
by using the preceding result by Yedla {\it et al.} \cite{Yedla}.
In addition, we compare it 
with the SD scheme of regular LDPC codes of SIC BP decoder when $\theta=0$ or 
$\pi/4$.
These studies show that our method outperforms 
the SD scheme with the LDPC codes
when the powers of both signals are not the same but are close to each other.

The work is organized as follows. 
Section II formulates our CAF scheme as well as the SD scheme,
and prepares several notions for our LDPC codes with CAF scheme.
Section \ref{random} discusses the SIR as the function of the angle difference $\theta$ between 
the constellation points.
Section \ref{S4} numerically compares asymptotic decodable regions
of the CAF scheme based on a degraded channel
and of the SD scheme with joint decoding
in both cases (the case of random coding and the case of LDPC codes).
Further, it compares our scheme with 
the SD scheme with the SIC BP decoding of LDPC codes.

\section{Preliminaries}

\begin{figure}[t]
    \centering
    \includegraphics[width=0.83\hsize]{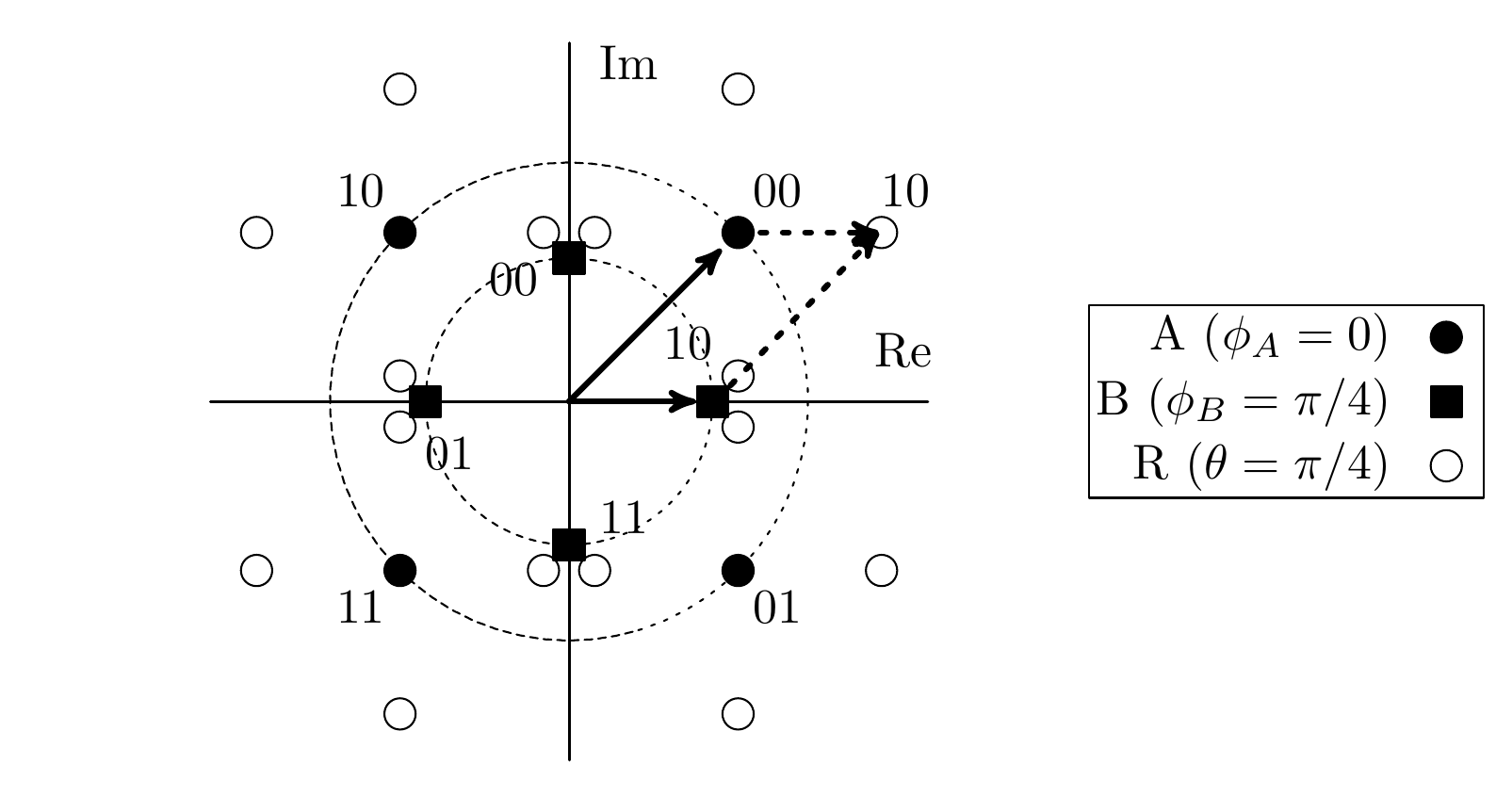}
    \caption{QPSK modulations at terminal A ($h_A=1$, $\phi_A=0$; filled circles) and B
     ($h_B=0.6$, $\phi_B=\pi/4$; filled squares) 
    and received signal points at relay R (open circles) in the noiseless case.
    Each label shows a bit sequence corresponding to the signal point. 
    As an example, when $(X_A^{(2t-1)},X_A^{(2t)})=(0,0)$ and $(X_B^{(2t-1)},X_B^{(2t)})=(1,0)$, 
    dotted arrows indicates the location of the received signal point whose label is $10$. 
    }
    \label{fig:conste}
\end{figure}

\subsection{Signal models}
We model the uplink of two terminals A and B to a relay or access point R by using two transmission phases.
In both scenarios, we employ 
the QPSK modulation, in which, possible values of input signals are 
limited to $\frac{e^{{j}\phi}}{\sqrt{2}} (\pm 1 \pm j)$, 
where {rotation angle $\phi$ of the constellation} is fixed for each transmitter.
Then, we divide the input signal into the real and imaginary parts,
that is, $n$ transmissions with complex signal 
can be converted to $2n$ transmissions with real signal. 
In the first phase,  
the following MAC channel is available.
Let $(X_A^{(2t-1)},X_A^{(2t)})$ (resp. 
$(X_B^{(2t-1)},X_B^{(2t)})$) be a pair of binary random variables with time index $t$.
Using the binary-bipolar conversion function $\mu: \{0,1\} \rightarrow \{+1, -1\}$ with $\mu(x) = 1-2 x$,
we apply the conversion 
{$\mu_\phi(x_1,x_2)= \frac{e^{{j}\phi}}{\sqrt{2}} (\mu(x_1)+j \mu(x_2))$ to
 $(X_A^{(2t-1)},X_A^{(2t)})$ with $\phi=\phi_A$ (resp. $(X_B^{(2t-1)},X_B^{(2t)})$ with $\phi=\phi_B$)}
  before their transmission {as the QPSK modulation.
The parameter $\theta := \phi_A- \phi_B$ plays a key role in decoding performance.
}
The terminals A and B then transmit the modulated signals 
$\mu_{\phi_A}(X_A^{(2t-1)},X_A^{(2t)})$ and 
$\mu_{\phi_B}(X_B^{(2t-1)},X_B^{(2t)})$ to the air. 
The relay R observes 
a received signal 
\begin{align}
	Y^{(t)} &= h_A\mu_{\phi_A}(X_A^{(2t-1)},X_A^{(2t)}) \nonumber\\
	&+ h_B\mu_{\phi_B}(X_B^{(2t-1)},X_B^{(2t)})+ W^{(t)},  \label{channel_model}
\end{align}
where $W^{(t)}$ is a zero mean complex Gaussian random variable with variance $\sigma^2$
and $h_A, h_B\ge 0$ respectively represent reception power of terminals A and B.
It is natural that they have different values. For simplicity, we assume that $h_A\ge h_B$.
{Figure~\ref{fig:conste} shows an example of constellations at two terminals and
the received signal points at the relay R in the noiseless case.}
In the second phase, the relay R transmits a certain information
to the respective terminals A and B by using the conventional QPSK Gaussian channel.
We discuss two schemes.
The first scheme {called CAF scheme} is given as follows.
In the first phase, 
both terminals A and B encode their messages ${\bf M}_A$ and ${\bf M}_B$ in 
$\{0,1\}^\ell$ to $\{0,1\}^{2n}$
by using the same linear code.
After receiving the symbol $Y^{(t)}$ via the channel 
\eqref{channel_model},  
the relay R infers ${\bf M}_A \oplus {\bf M}_B \in \{0,1\}^\ell$
as correct as possible, 
where the operator $\oplus$ represents the addition over $\F_2$.  
This phase is called the MAC phase.
The channel is ideal if $h_A = h_B = 1$
though it is not the case in practice.

In the second phase {of the CAF scheme}, the relay R broadcasts the estimate of 
${\bf M}_A \!\oplus\! {\bf M}_B $ to both terminals.
If the estimate equals to 
the true value ${\bf M}_A \oplus {\bf M}_B $, 
the terminal A (resp. B) can retrieve ${\bf M}_B$ (resp. ${\bf M}_A$) 
from ${\bf M}_A \oplus {\bf M}_B $
and original ${\bf M}_A$ (resp. ${\bf M}_B$).
This phase is called the broadcast phase.
This scheme, i.e., wireless network coding \cite{Katti08} \cite{Narayanan},  can be seen as the simplest case of 
the CAF technique \cite{Nazer11}.
Since the broadcast phase is simple transmission,
it has larger capacity than the MAC phase.
We thus focus only on the MAC phase as 
it is the bottleneck of this scheme.

In the second scheme called the SD scheme, 
at the first phase,
simultaneously, both terminals A and B encode their messages ${\bf M}_A$ and ${\bf M}_B$ 
in $\{0,1\}^{\ell}$ 
by using different linear codes $f_A$ and $f_B$,
which map ${\bf M}_A$ and ${\bf M}_B$ into $\{0,1\}^{2n}$, respectively.
Here, the rotation angles
$\phi_A$ and $\phi_B$ are set to different values to maximize the transmission rate.
Then, the relay R decodes both messages ${\bf M}_A$ and ${\bf M}_B$
from the superimposed signals.
The first phase is the same as the conventional MAC coding with the same transmission rate.
In the second phase, the relay R sends ${\bf M}_A$ and ${\bf M}_B$ to the respective terminals.
Since the second phase is simple transmission,
it has larger capacity than the first phase.
Hence, we focus only on the first phase
even in the SD scheme.

\subsection{Degraded Channel Model}\label{2B}
In the CAF scheme, 
the two terminals A and B are assumed to use the same code
$C \subset \{0,1\}^{2n}$ for their encoding.
To infer the modulo sum ${\bf M}_A \oplus {\bf M}_B $
by a BP decoder,
we use the symbol log-likelihood ratio (LLR).
However, the BP decoder has too large calculation complexity
when it is designed from the LLR based on the true channel
because the LLR has a complicated form even for LDPC codes.
To resolve this problem, in the decoding step, 
we employ the following degraded channel instead of the real channel. 

We now consider a virtual channel whose input and output symbols are $Z^{(t)}$ and $Y^{(t)}$, respectively.
When $Z^{(t')} = 0$, 
we set $X_A^{(t')} = X_B^{(t')} = 0$ with probability $1/2$,
and 
set $X_A^{(t')} = X_B^{(t')} = 1$ with probability $1/2$ for $t'=1, \ldots, 2n$.
In addition, when $Z^{(t')} = 1$, 
we set the remaining two cases with probability $1/2$.

In the remaining of this subsection, we set the angle parameters
$\phi_A$ and $\phi_B$ to the same value $0$.
The reason for this setting will be explained in Section \ref{random}.
In this case, we can convert the QPSK scheme to the BPSK scheme.
For this conversion, we introduce real random variables $Y_R^{(2t-1)}$ and $Y_R^{(2t)}$ as
$Y^{(t)}=(Y_R^{(2t-1)}+j Y_R^{(2t)})/\sqrt{2}$,
which satisfies
\begin{align}
	Y_R^{(t')} &= h_A\mu(X_A^{(t')}) 
+ h_B\mu(X_B^{(t')})+ W^{(t')}_R,  \label{channel_model2}
\end{align}
where $W^{(t')}_R$ is a zero mean complex Gaussian random variable with variance $\sigma^2$.
Since $Z^{(t')}\!=\!X_A^{(t')}\!\oplus\! X_B^{(t')}$,
the conditional PDF representing the channel statistics 
is 
\begin{equation} 
\begin{aligned}
\!\mathrm{Pr} [Y_R^{(t')}\! =\! y | Z^{(t')}\! =\! x]\! &= 
\frac{1}{2} F_R(y; h_A+ (-1)^x h_B, \sigma^2) \\
&+\!\frac{1}{2} F_R(y; -h_A\!-\!(-1)^x h_B, \sigma^2), 
\end{aligned}\label{vir}
\end{equation}
where $F_R(y; m, \sigma^2)$ is the Gaussian distribution with mean $m$ and variance $\sigma^2$
over real numbers.
This channel \eqref{vir} is called the degraded channel 
\cite{Ullah17}.


The symbol LLR function is derived as in~\cite{Sula17}:
\begin{equation} \label{llr}
\lambda^{({t'})}(y) 
= \ln \left[ \frac{\cosh \frac{y(h_A+h_B)}{\sigma^2}}{\cosh \frac{y(h_A-h_B)}{\sigma^2}} \right]-\frac{2h_Ah_B}{\sigma^2}.
\end{equation}

Let us turn to the argument on the case where terminals A and B 
employ a binary linear code $C$.
Since $(X_A^{(1)}, \ldots, X_A^{(2n)})$
and $(X_B^{(1)}, \ldots, X_B^{(2n)})$ belong to $C$, 
the linearity of the code $C$ guarantees that
$(Z^{(1)},\ldots, Z^{(2n)})$ also belongs to $C$.
From this fact, {\em degraded channel based ML
decoding} can be defined as
\begin{equation} \label{iidml}
(\hat z_1, \ldots, \hat z_{2n}) \!=\! \argmax_{(z_1,\ldots, z_{2n}) \in C}
\prod_{t' = 1}^{2n} 
\mathrm{Pr} [Y_R^{(t')} \!=\! y_{t'} | Z^{(t')} \!=\! z_{t'}].
\end{equation}
Here, the BP decoder is aimed to approximately solve 
the maximization \eqref{iidml}.
Though this decoding rule is different from the true ML decoder for $Z$, 
this decoding rule significantly reduces the amount of the calculation 
in the decoding process, and is employed for decoding in 
the CAF scheme.

\subsection{LDPC coding}
Due to the discussion in Section \ref{random}, we set 
the parameter $\theta=0$ 
 in the CAF scheme.
Let $C $ be an LDPC code used in terminals A and B. The terminals A and B 
independently select own codewords 
$\mathbf{x}_A = (x_{A,1},\ldots, x_{A, 2n})  \in C$
and $\mathbf{x}_B = (x_{B,1},\ldots, x_{B, 2n}) \in C$ 
according to their own message. 
Since 
{$Y^{(t)}=(Y_R^{(2t-1)}+j Y_R^{(2t)})/\sqrt{2}$, 
based on the channel model (\ref{channel_model}) with $\theta=0$,}
the received real signal $\mathbf{y}_R =
(y_R^{(1)}, \ldots, y_R^{(2n)} )$ can be converted as
\begin{equation}
\begin{aligned}
	\mathbf{y}_R 
	=&   h_A (\mu(x_{A,1}),\ldots, \mu(x_{A, 2n})) \\
        &+  h_B (\mu(x_{B,1}),\ldots, \mu(x_{B, 2n})) 
	+ \mathbf{w}_R,
	\label{eq_caf1}
\end{aligned}
\end{equation}
where $\mathbf{w}_R$ is an additive real Gaussian noise vector.
A decoder like a BP decoder  tries to recover 
$\mathbf{x}_A \oplus \mathbf{x}_B$ from the received word $\mathbf{y}_R$.
Our goal is to evaluate decoding performance of LDPC codes
 to recover $\mathbf{x}_A \oplus \mathbf{x}_B$ even when $h_A\neq h_B$.

In the decoding of LDPC codes,
it is very hard to implement the {BP decoder based on the true ML decoding
as discussed in the last subsection}.
Since BP decoding for LDPC codes can be regarded as an 
approximation of ML decoding, 
we employ the BP decoder based on the degraded channel, which equals  
the conventional log-domain {binary} BP algorithm \cite{Richardson} with the symbol LLR expression (\ref{llr}).

This type of BP decoding has already discussed in \cite{Sula17} 
\cite{Takabe18}.
Its notable advantage is that it can be easily implemented based on 
a practical BP decoder for the additive white Gaussian noise (AWGN) channel just by 
replacing an LLR computation unit.

Yedla {\it et al.} \cite{Yedla} also discussed 
joint BP decoding for LDPC codes with the SD scheme in the BPSK modulation.
Since two transmissions on the BPSK modulation can be regarded as
a single transmission on the QPSK modulation with $\phi_A=\phi_B$,
their results are applicable to the SD scheme with the QPSK modulation when $\theta =0$. 
However, it is not so easy to implement the joint BP decoding in the SD scheme
 in practice due to the following two reasons.
First, while the BP decoder is constructed from the corresponding factor graph, 
the factor graph of the joint BP decoder 
is composed of two factor graphs of respective LDPC codes, and is so large that the circuit of the joint BP decoder requires large power.
Second, since existing hardware cannot be used for the joint BP decoder,
it requires new hardware implementation
while the BP decoder for the CAF scheme based on the degraded channel
can be implemented by using existing hardware.
In this sense, our BP decoding for the CAF scheme is practical.

{Another decoding scheme in the SD scheme is SIC BP decoding.}
In the first type of SIC BP decoding,
in order to decode ${\bf M}_A$,
we apply a BP decoder based on the channel \eqref{vir2} to the received signal 
$\mathbf{y}$. 
{In this decoding step, similar to (\ref{eq_caf1}), 
we can use a binary BP decoder with an LLR function
defined by a channel model,
\begin{equation}
	\mathbf{y}_R 	= h_A (\mu(x_{A,1}),\ldots, \mu(x_{A, 2n}))+ \mathbf{w}'_R,
\label{eq_sic0}
\end{equation}  
where $\mathbf{w}'_R$ is an i.i.d. random vector whose distribution is 
\begin{equation}
{\frac{1}{2}F_R\left(w'_R; {h_B}, {\sigma^2}\right)
+\frac{1}{2}F_R\left(w'_R; -{h_B},{\sigma^2}\right)},
\label{eq_sic1}
\end{equation}  
when {the difference of rotation angles} is $\theta =0$, and 
\begin{align}
\frac{1}{4}&F_R\left(w'_R; {\sqrt{2}}h_B, \sigma^2\right)
+\frac{1}{2}F_R\left(w'_R; 0, \sigma^2\right)\nonumber\\
&+\frac{1}{4}F_R\left(w'_R; -{\sqrt{2}}{h_B}, \sigma^2\right),
\label{eq_sic2}
\end{align}
when $\theta = \pi/4$.
}  
Then, in order to decode ${\bf M}_B$,
we apply a BP decoder 
to $\mathbf{y} ':=\mathbf{y} -h_A \mu_{\phi_A} ({\bf M}_A)$.
In the second type of SIC BP decoding,
we exchange the roles of ${\bf M}_A$ and ${\bf M}_B$.
While this method has no {implementation issues} unlike joint BP decoding,
it works only near the corner points of the capacity region \cite{Palanki}.
This problem can be resolved by employing time sharing \cite{Rimoldi}.
However, we cannot freely choose the coding rate pair when we employ LDPC codes.
Due to this restriction, only time sharing of very limited class of rate pairs
is available.
For simplicity, in this paper, we address LDPC codes with SIC BP decoding
without time sharing.


\section{Linear Random Coding}\label{random}
First, for the CAF scheme,
we employ the linear random codes for the choice of $C $
and the degraded channel based ML decoding.
Then, as the single use of the channel,
given two pairs of binary variables $(X_{A,R},X_{A,I})$ and 
$(X_{B,R},X_{B,I})$ subject to the uniform distribution independently,
we consider the channel
\begin{equation} \label{channel_model1}
	Y= h_A\mu_{\phi_A}(X_{A,R},X_{A,I}) 
	+ h_B\mu_{\phi_B}(X_{B,R},X_{B,I})+ W, 
\end{equation}
where $W$ is a zero mean complex Gaussian random variable with variance $\sigma^2$.
Based on \eqref{channel_model1},
we focus on 
the mutual information 
\begin{equation} \label{mutual1}
 I(Y_R; X_{A,R}+X_{B,R},X_{A,I}+X_{B,I}),
\end{equation}
when 
the variables 
$X_{A,R}$, $X_{A,I}$, $X_{B,R}$, and $X_{B,I}$ are
subject to the uniform distribution independently.
Due to this symmetric assumption for the distribution of the input,
the value \eqref{mutual1} is called SIR in the CAF scheme.
Then, in the CAF scheme,
the mutual information rate $ I(Y_R; X_{A,R}\!+\!X_{B,R},X_{A,I}\!+\!X_{B,I})$ 
is achievable by LRC\cite{Ullah17}.
Since this value depends on the difference 
$\theta=\phi_A-\phi_B$
between two angle parameters $\phi_A$ and $\phi_B$, we need to choose a suitable parameter $\theta$.

In the SD scheme, 
when the rates of ${\bf M}_A$ and ${\bf M}_A$ 
are given as $R_1$ and $R_2$,
the achievable rate by LRC is given as
\begin{align*}
\{ (R_1,R_2)| &R_1 \le I(Y;X_{A,R},X_{A,I}| X_{B,R},X_{B,I}),\nonumber\\  
&R_2 \le I(Y;X_{B,R},X_{B,I}| X_{A,R},X_{A,I}),\nonumber\\
& R_1 +R_2 \le I(Y;X_{A,R},X_{A,I},X_{B,R},X_{B,I})\},
\end{align*}
when we employ the joint ML decoding\cite{Ahlswede,Liao}.
Here, the variables 
$X_{A,R}$, $X_{A,I}$, $X_{B,R}$, and $X_{B,I}$ are assumed to be
subject to the uniform distribution independently.
In this notion, the variables $X_A$ and $X_B$ are independently subject to the uniform distribution on 
$\F_2$.
{If we assume that the rate of ${\bf M}_A$ is equal to that of ${\bf M}_B$, the rate}
\begin{align}
\min(&I(Y; X_{A,R},X_{A,I},X_{B,R},X_{B,I})/2, \nonumber \\
& I(Y;X_{A,R},X_{A,I}|X_{B,R},X_{B,I}),  \nonumber \\
& I(Y;X_{B,R},X_{B,I}|X_{A,R},X_{A,I}))
\label{LHY}
\end{align}
is achievable as the rate of ${\bf M}_A$.
Due to the same reason as \eqref{mutual1},
the value \eqref{LHY} is called the SIR in the SD scheme.
Since the value given in \eqref{LHY} depends on the parameter 
$\theta=\phi_A-\phi_B$,
we need to choose a suitable parameter $\theta$.

{Instead of joint ML decoding,} there are two types of SIC decoders
{in the SD scheme}. 
In the first type, we first decode ${\bf M}_A$ 
based on the channel from the received signal $\mathbf{y} $;
\begin{equation} 
\begin{aligned}
&\mathrm{Pr} [Y^{(t)} = y | X_A^{(2t-1)} = x_1, X_A^{(2t)} = x_2] \\
=& 
\sum_{(x_1',x_2') \in \{0,1\}^2}
\frac{1}{4} 
F(y;h_A \mu_{\phi_A}(x_1,x_2) + h_B \mu_{\phi_B}(x_1',x_2') , \sigma^2),
\end{aligned}
\label{vir2}
\end{equation}
{where $F(y;m,\sigma^2)$ is the complex Gaussian distribution with mean $m$ and 
variance $\sigma^2$.} 
Then, we decode ${\bf M}_B$ by applying the decoder to 
$\mathbf{y} ':=\mathbf{y}-h_A \mu_{\phi_A} ({\bf M}_A)$
based on the channel; 
\begin{align} 
& \mathrm{Pr} [{Y'}^{(t)} = y |X_B^{(2t-1)} = x_1', X_B^{(2t)} = x_2'] \nonumber \\
=&F(y; h_B \mu_{\phi_B}(x_1',x_2'), \sigma^2) .
\label{vir3}
\end{align}
In this case, the rate
\begin{equation} 
\min(I(Y;X_{A,R},X_{A,I}), \allowbreak I(Y;X_{B,R},X_{B,I}| X_{A,R},X_{A,I}))
\nonumber
\end{equation} 
 is achievable \cite{Palanki}.
In the second type, the role of ${\bf M}_A$ is exchanged by that of ${\bf M}_B$. 
Hence, the rate 
\begin{equation} 
\min(I(Y;X_{A,R},X_{A,I}| X_{B,R},X_{B,I}),I(Y;X_{B,R},X_{B,I}))
\nonumber
\end{equation} 
 is achievable.
Time sharing enables us to achieve the rate
\cite{Rimoldi};
\begin{align} 
\max_{0\le \lambda\le 1}
\min\Big(&\lambda I(Y;X_{A,R},X_{A,I})\nonumber\\
&\,+(1-\lambda)I(Y;X_{A,R},X_{A,I}| X_{B,R},X_{B,I}),\nonumber\\
&\lambda I(Y;X_{B,R},X_{B,I}| X_{A,R},X_{A,I})\nonumber\\
&\,+(1-\lambda)
I(Y;X_{B,R},X_{B,I})\Big),
\label{vir20}
\end{align}
which equals to \eqref{LHY}.

\begin{figure}[t]
    \centering
    \includegraphics[width=0.85\hsize]{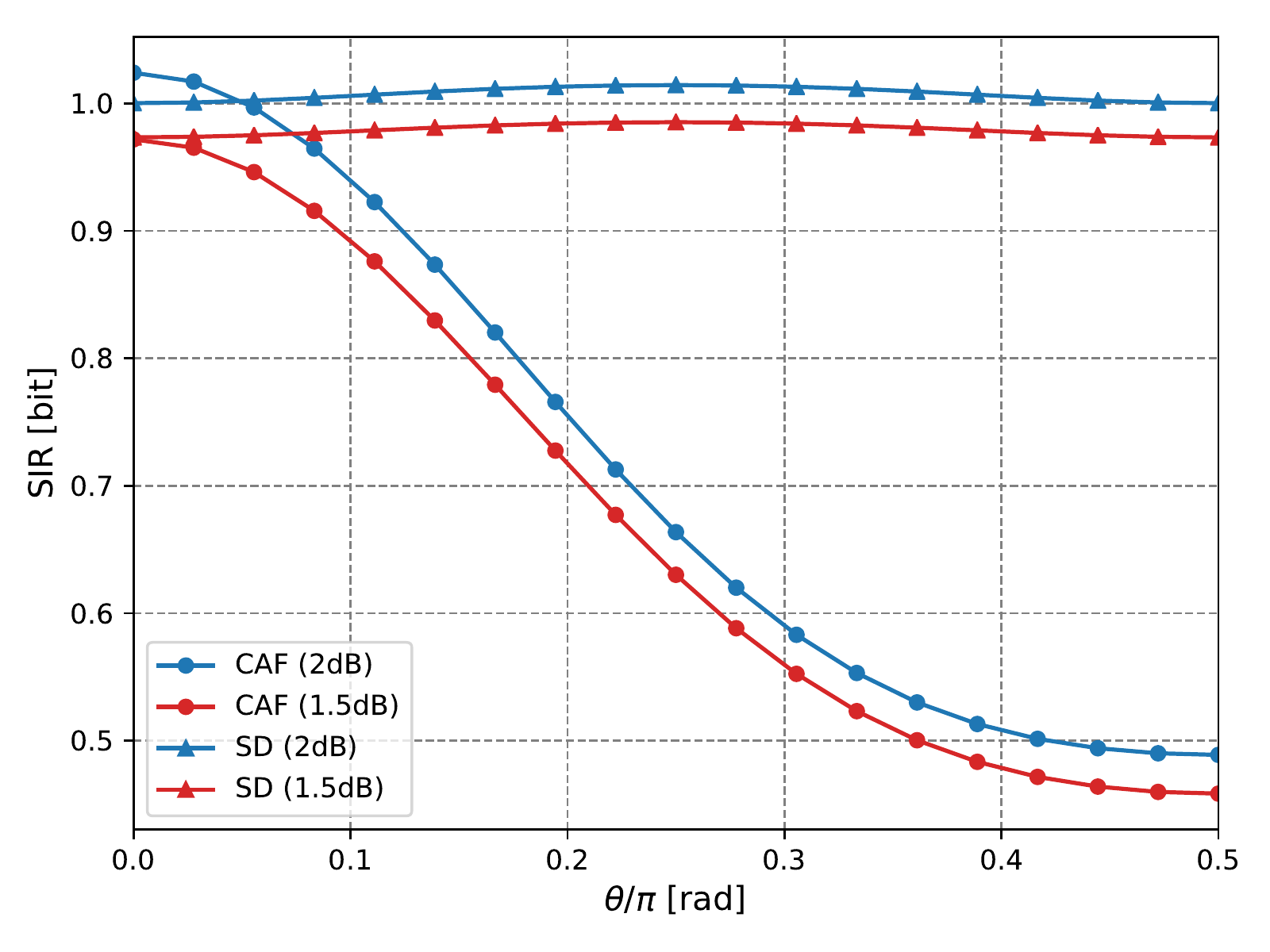}
    \caption{{Angle} difference $\theta$ dependency of the SIR of the CAF (circles) 
    and SD (triangles) scheme.
    The SNR of terminal A is fixed to $2$dB.}
    \label{fig:phasediff}
\end{figure}

{To examine the dependency} of the parameter
$\theta=\phi_A-\phi_B$ between the two rotation angles,
Fig.\ref{fig:phasediff} 
shows SIRs given in 
\eqref{mutual1} and \eqref{LHY} of both schemes.
Here, signal-to-noise ratios (SNRs) of two terminals are respectively defined by
$\mbox{SNR}_A = 10 \log_{10} (h_A^2/\sigma^2)$ [dB]
and $\mbox{SNR}_B = 10 \log_{10} (h_B^2/\sigma^2)$ [dB].
In Fig.\ref{fig:phasediff}, we fix SNR$_A$ to $2$dB ($\sigma=0.7943$, $h_A=1$) and
SNR$_B$ is set to $2$ and $1.5$dB, which respectively correspond to $h_B=1$, and $0.944$.
Note that the preceding study \cite{Mu} only discussed the case where $h_A=h_B$.
In the CAF scheme,
the maximum SIR is realized when 
 $\theta$ equals to zero.
In particular, the maximum value is much larger than the value in other cases.
This numerical analysis suggests us to fix the parameter $\theta$ to zero in the CAF scheme
as we set in Subsection \ref{2B}.
Oppositely,
in the SD scheme, the maximum SIR is realized when the difference $\theta$ equals ${\pi}/{4}$.
However, the difference between the maximum and other values is not so large.
This fact suggests us that 
the optimal choice of the parameter $\theta$ might depend on the choice of our code.

\begin{figure}[t]
    \centering
    \includegraphics[width=0.995\hsize]{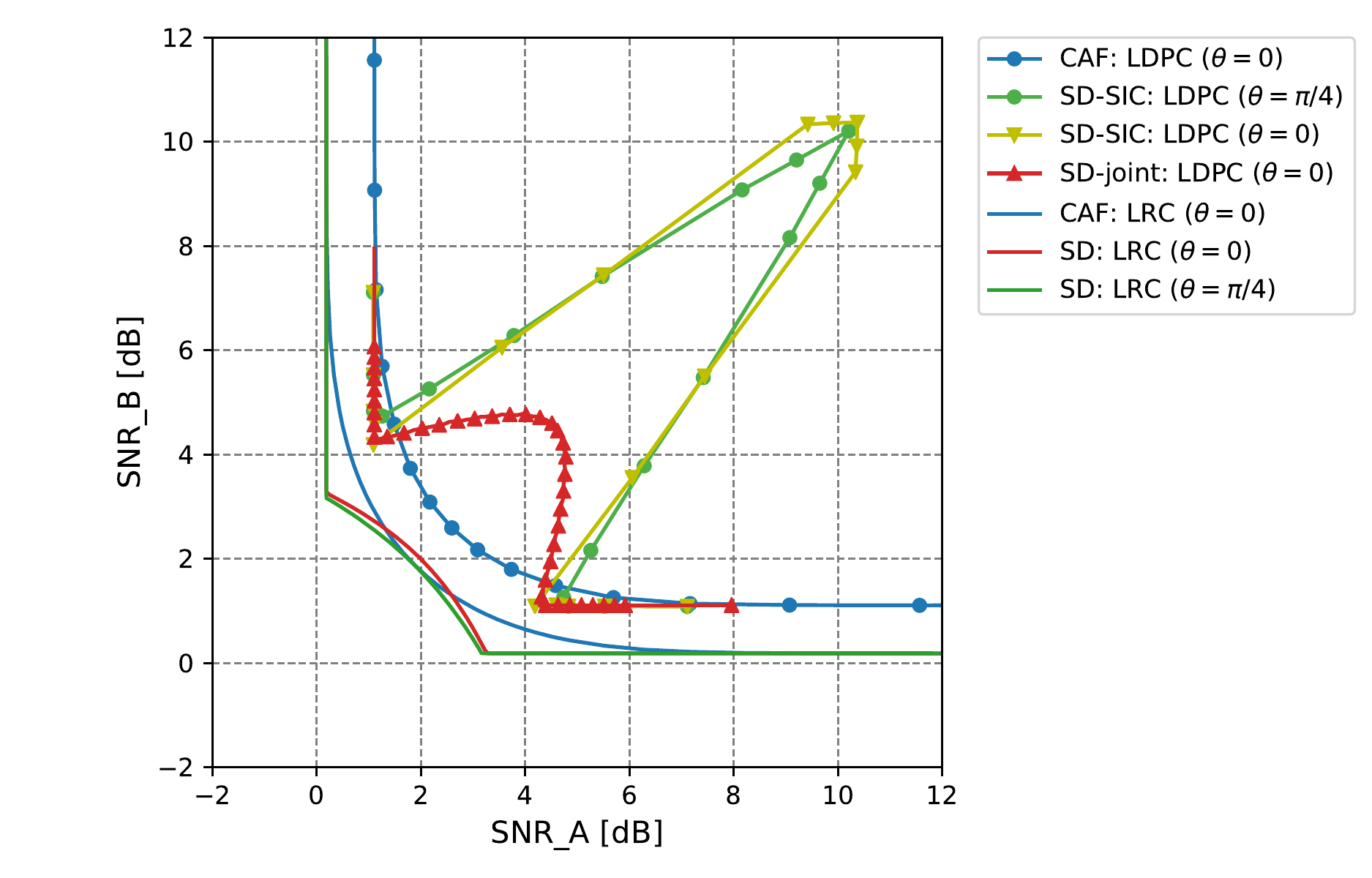}
    \caption{ACPR when the code rate is $1/2$. 
   {Solid lines represent ACPRs of the LRC evaluated by the SIR of the CAF scheme (blue; $\theta=0$) and SD scheme with $\theta=0$ (red) and $\theta=\pi/4$ (green).
    Symbols represents the asymptotic decodable region of $(3,6)$-LDPC does for the CAF scheme (blue), SD scheme with a joint BP decoder~\cite{Yedla} when $\theta=0$ (red)
    and with a SIC BP decoder when $\theta=0$ (yellow) and $\theta=\pi/4$ (green).}}
    \label{fig:rate_1_2_delta_bp}
\end{figure}

\section{Asymptotic decodable region}\label{S4}
{
In this section, the CAF scheme based on the degraded channel
is compared with 
the SD scheme with the joint decoding and with the SIC decoding
in terms of the asymptotic channel parameter region (ACPR)~\cite{Yedla}
 with respect to SNR$_A$ and  SNR$_B$.
The ACPR represents a decodable channel parameter region for a given rate.   
To obtain the ACPR of  LDPC coding for the degraded CAF scheme, we 
calculate the asymptotic decodable region by the density evolution technique
whose detail procedure is explained in \cite{Takabe18}.
On the other hand, Yedla {\it et al.} \cite{Yedla} derived 
the ACPR for LDPC codes with the SD scheme of a joint BP decoder 
with another coordinate $(h_A, h_B)$ under the BPSK modulation.
We rewrite their graph with our coordinate $(\mbox{SNR}_A,\mbox{SNR}_B)$
by converting the BPSK modulation to the QPSK modulation
with $\theta=0$.
The other  ACPR for LDPC coding with the SD scheme of a SIC BP decoder
is evaluated by conventional density evolution technique using LLR functions corresponding to
(\ref{eq_sic1}) and (\ref{eq_sic2}).   
As described in the last section, we find that 
the difference between the two rotation angles $\phi_A$ and $\phi_B$ is crucial 
to maximize the SIR. 
We thus set the parameter $\theta$ to zero in the CAF scheme
whereas we consider two cases $\theta = 0,\pi/4$ in the SD scheme whose
ACPR will depend on the combination of a coding scheme and the choice of $\theta$.
}

Fig.~\ref{fig:rate_1_2_delta_bp} shows
the ACPRs of $(3,6)$-regular LDPC codes for each scheme and those of LRC
obtained by the SIR when the coding rate is $1/2$.
{In this case, the ACPR of  LRC in the SD scheme with $\theta= \pi/4$ is wider than that in the SD scheme indicating that LRC in the SD scheme achieves better performance by time sharing. }
On the other hand, we found that the $(3,6)$-regular LDPC code with the CAF scheme
has better performance than that with the SD scheme when 
{$\mbox{SNR}_A$ and $\mbox{SNR}_B$ are smaller than $4$dB, i.e., 
$0.7h_A\le h_B\le 1.3h_A$.
In this region, the SIC BP decoder of the $(3,6)$-regular LDPC code for the SD scheme
shows a poor performance because the channel~(\ref{eq_sic0}) has a relatively large noise due to 
a transmitted signal from {the other} terminal.
In addition, even if a computationally expensive joint BP decoder is used in the SD scheme, 
its ACPR is narrower than that of the CAF scheme when $h_A$ is close to $h_B$.
Although the $(3,6)$-regular LDPC code in the SD scheme shows better performance 
at corner points of the ACPR, it is suggested that LDPC coding with the CAF scheme 
is better than that with the SD scheme except for the limited parameter region when the rate is $1/2$.
}

\begin{figure}[t]
    \centering
    \includegraphics[width=0.995\hsize]{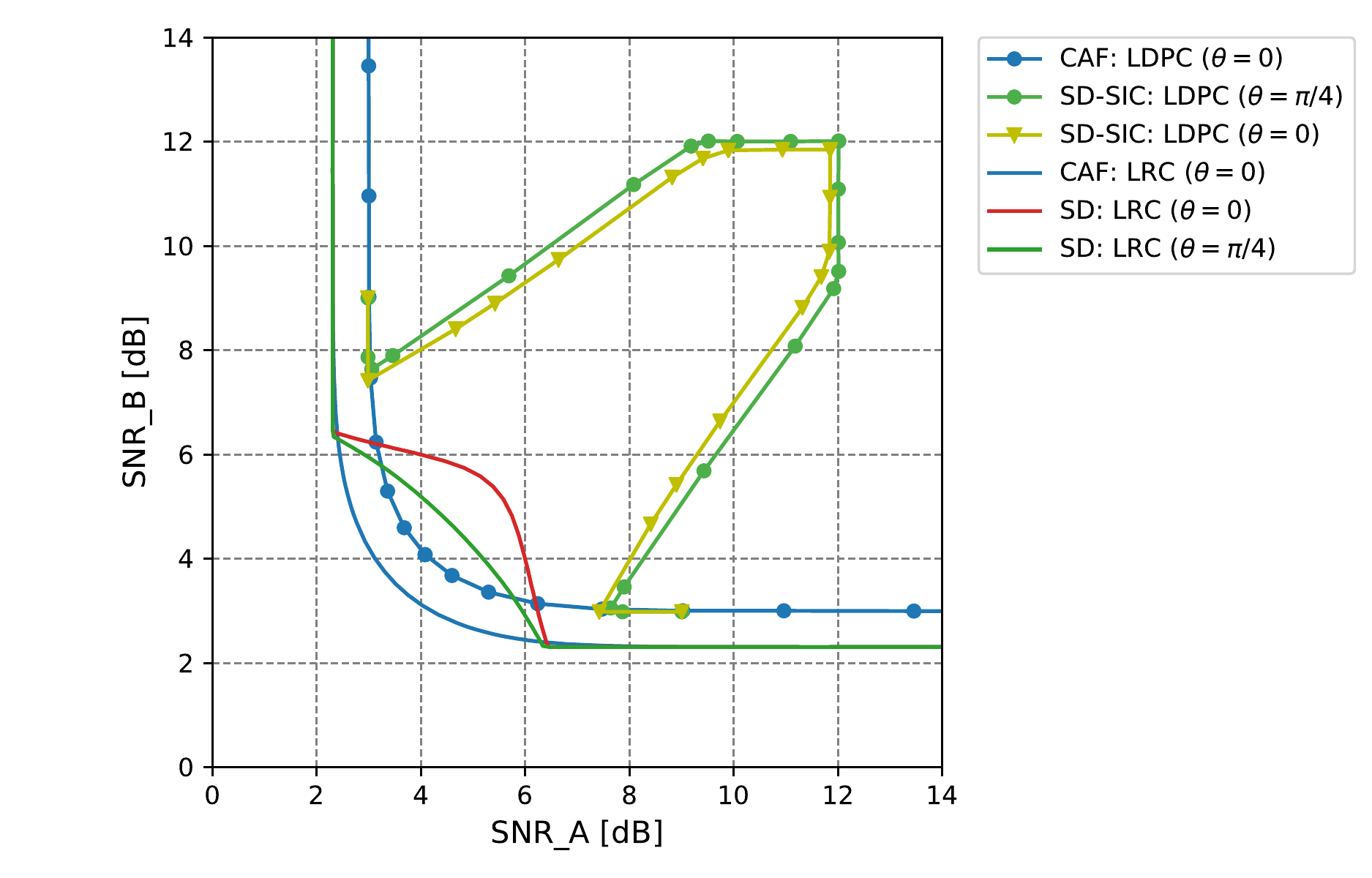}
    \caption{ACPR when the code rate is $2/3$. 
       {Solid lines represent ACPRs of the LRC evaluated by the SIR of the CAF scheme (blue; $\theta=0$) and SD scheme with $\theta=0$ (red) and $\theta=\pi/4$ (green).
    Symbols represents the asymptotic decodable region of $(3,9)$-LDPC does for the CAF scheme (blue)
     and the SD scheme with SIC BP decoding when $\theta=0$ (yellow) and $\theta=\pi/4$ (green).}}
    \label{fig:rate_1_3_delta_bp}
\end{figure}

Fig.~\ref{fig:rate_1_3_delta_bp} shows 
{ACPRs of LDPC and LRC coding with the CAF and SD schemes when the coding rate is $2/3$.
In this coding rate, the LRC in the CAF scheme 
has a wider ACPR than that in the SD scheme except for the very limited parameter region.
It suggests that the ACPR of the LDPC codes with the joint BP decoder in the SD scheme
is narrower than that in the CAF scheme when $h_A$ is sufficiently close to $h_B$.
This is because the ACPR of LRC is the outer bound of the  ACPR of  LDPC coding.
Moreover, we found that the the $(3,9)$-regular LDPC codes with the CAF scheme 
shows better performance than the LRC with the SD scheme
if $0.8h_A\le h_B\le 1.2h_A$.
The gap of ACPRs becomes considerably large when we compare it with the LDPC coding with the SD scheme
using the SIC BP decoder.  }
Hence, in a NOMA scenario with a wide range of SNRs, 
our proposed scheme based on the CAF scheme has better performance than 
the SD scheme. 

In contrast, for the coding rate $1/3$,
we found that the LRC with the SD scheme with the joint ML decoder 
is better than
the LRC with the degraded CAF scheme with the ML decoder 
although we do not present the graph of such a case due to the page limitation.
This fact suggests that 
our method is not so good as the SD scheme 
for the rate {below} $1/3$.


\section{Concluding remarks}
In this paper, we have investigated the behavior of the LDPC coding 
for CAF relaying, which is useful for wireless two-way relay channels.
{We theoretically analyzed the decoding performance 
of LRC and LDPC codes in the CAF and SD schemes, which depends on 
the difference of rotation angles of constellations.} 
The results show that our scheme has advantages over the SD schemes
 with SIC BP or joint BP decoding in the high rate region.
Therefore, our method {is potentially useful} for NOMA in 
the power domain such as integrated wireless networks
 to boost throughput while ensuring fair rate allocation in cases of unbalanced SNRs.
Further, our constructed code can be used for 
 secure communication via untrusted relay~\cite{MWV}.

\subsection*{{\bf Acknowledgement}}
This study is partly supported by JSPS KAKENHI Grant-in-Aid for Scientific Research (A) 17H01280.


\begin{thebibliography}{99} 

\bibitem{Miridakis}
N. I. Miridakis and D. D. Vergados
``A Survey on the successive interference cancellation performance for single-antenna and multiple-antenna OFDM systems,''
{\em IEEE Communications Surveys \& Tutorials}, vo. 15, no. 1, pp.  312-335, 2013.

\bibitem{Palanki}
R. Palanki, A. Khandekar, and R. J. McEliece, 
``Graph-based codes for synchronous multiple access channels,''
\textit{Proc. 31st Ann. Allerton Conf. Commun., Contr. Comput.}, Monticello, IL, 2001, pp. 1263-1271.

\bibitem{Rimoldi}
B. Rimoldi and R. Urbanke, 
``A rate-splitting approach to the Gaussian multiple-access channel,'' 
{\em IEEE Trans. Inf. Theory}, vol. 42, no. 2, pp. 364-375, 1996.

\bibitem{Yedla}
A. Yedla, P. S. Nguyen, H. D. Pfister, and K. R. Narayanan,
``Universal codes for the gaussian MAC via spatial coupling,''
\textit{2011 49th Annual Allerton Conf. Commun., Control, Comp.}, Monticello, IL, 2011, pp. 1801-1808.


\bibitem{Nazer11} 
B. Nazer and M. Gastpar,
``Compute-and-forward: harnessing interference through 
structured codes,'' 
{\em IEEE Trans. Inf. Theory}, vol. 57, no. 10, pp. 6463-6486, 2011.

\bibitem{Lu14}
L. Lu, L. You, and S. C. Liew, 
``Network-coded multiple access,'' 
{\em IEEE Trans. Mobile Computing}, 
vol. 13, no. 12, pp. 2853-2869, 2014. 

\bibitem{Guo2018}
C. Guo {\em et Al.}, 
``Compute-and-forward for uplink non-orthogonal multiple access,'' 
IEEE Wireless Comm. Lett., vol. 7, no. 6, pp. 986-989, Dec. 2018. 

\bibitem{Nazer2016} 
B. Nazer, V. Cadambe, V. Ntranos and G. Caire, 
``Expanding the compute-and-forward framework: Unequal powers, signal levels, and multiple linear combinations,''
{\em IEEE Trans. Inf. Theory}, vol. 62, no. 9, pp. 4879-4909, 2016.

\bibitem{MacKay99}
D. J. C. MacKay, 
``Good error correcting codes based on very sparse matrices, ''
{\em IEEE Trans. Inf. Theory}, vol. 45, no. 2, pp. 399-431, 1999.


\bibitem{Sula17}
E. Sula,  J. Zhu,  A. Pastore,  S. H.Lim,  and  M. Gastpar, 
``Compute-forward multiple access (CFMA) with nested LDPC codes, ''
\textit{Proc. IEEE Int. Symp. Inf. Theory (ISIT2017)}, Aachen, Germany, pp. 25-30 June 2017, pp. 2935-2939.

\bibitem{Takabe18}
S. Takabe, Y. Ishimatsu, T. Wadayama, and M. Hayashi,
``Asymptotic analysis on spatial coupling coding for two-way relay channels,''
\textit{Proc. IEEE Int. Symp. Inf. Theory (ISIT2018)}, 
CO, 17-22 June 2018, pp. 1021-1025.


\bibitem{Richardson}
T. Richardson and R. Urbanke, {\it Modern coding theory}, Cambridge University Press, 2008.

\bibitem{Ullah17}
S. S. Ullah, G. Liva, and S. C. Liew,
``Physical-layer network coding: a random coding error exponent perspective, ''
\textit{Proc. IEEE Inf. Theory Workshop (ITW)}, Kaohsiung, Taiwan, pp. 6-10 Nov. 2017,
pp.  559-563.


\bibitem{Katti08}
S. Katti,  H. Rahul,  W. Hu,  D.  Katabi,  M. Medard, and  J. Crowcroft,
``XORs in the air: practical wireless network coding, ''
{\em IEEE/ACM Trans. Networking}, vol. 16, no. 3, pp. 497-510, 2008.

\bibitem{Narayanan}
K. Narayanan, M. P. Wilson, and A. Sprintson, 
``Joint physical layer coding and network coding for bi-directional relaying, ''
\textit{Proc. 45th Ann. Allerton Conf. Commun., Contr. Comput.}, Monticello, IL, Sep. 2007, pp. 5641-5654.

\bibitem{Ahlswede}
R. Ahlswede, ``Multi-way communication channels,'' 
{\em Proc. 2nd Int. Symp. Inf. Theory (Thakadsor, Armenian SSR, Sep. 1971)}. 
Budapest, Hungary: Academia Kiado, 1971, pp. 23-52.

\bibitem{Liao}
H. Liao, 
{\em Multiple access channels}, Ph.D. dissertation, 
Dept. Electr. Eng., University of Hawaii, Honolulu, 1972.



\bibitem{Mu}
M. Wu, D. Wuebben, and  A. Dekorsy,
``Mutual information based analysis for physical-layer network coding with optimal phase control''
\textit{Proc. 9th International ITG Conf. Systems, Comm. and Coding},  M{u}nchen, Feb, 2013. 

\bibitem{MWV}
M. Hayashi, T. Wadayama, and A. Vazquez-Castro,
``Secure computation-and-forward communication with linear codes,''
{\em Proc. Inf. Theory Workshop (ITW2018)}, Guangzhou, 2018, pp. 1-5.



\end{thebibliography}
\end{document}